\documentclass[
  journal=pasa,
  manuscript=research-paper, %% or "review"
  year=2022,
  volume=37,
]{cup-journal}

\usepackage{microtype,siunitx,booktabs}
\usepackage{graphicx}	% Including figure files
\usepackage{amsmath}	% Advanced maths commands
\usepackage{amssymb}	% Extra maths symbols
\usepackage{rotating}
\usepackage{caption}
\usepackage{xcolor}
\usepackage{soul}

\newcommand{\kms}{\,km\,s$^{-1}$}

\newcommand{\around}{$\sim$}

\newcommand{\Msun}{M$_{\odot}$}

\newcommand{\Sersic}{S\'{e}rsic}

\title[Stellar Populations of Bulges and Discs]{Comparison of the Stellar Populations of Bulges and Discs using the MaNGA Survey}

\author{Philip Lah}
\affiliation{Research School of Astronomy and Astrophysics, Australian National University, Canberra, ACT 2611, Australia}
\email[Philip Lah]{philip.lah@anu.edu.au}

\author{Nicholas Scott}
\affiliation{Sydney Institute for Astronomy, School of Physics, University of Sydney, NSW 2006, Australia}
\alsoaffiliation{ARC Centre of Excellence for All Sky Astrophysics in 3 Dimensions (ASTRO 3D)}

\author{Tania M.~Barone}
\affiliation{Centre for Astrophysics and Supercomputing, Swinburne University of Technology, P.O. Box 218, Hawthorn, VIC 3122, Australia}
\alsoaffiliation{ARC Centre of Excellence for All Sky Astrophysics in 3 Dimensions (ASTRO 3D)}

\author{A.~S.~G.~Robotham}
\affiliation{International Centre for Radio Astronomy Research, University of Western Australia, 35 Stirling Highway, Crawley WA 6009, Australia}

\author{Francesco D'Eugenio}
\affiliation{Cavendish Laboratory and Kavli Institute for Cosmology, University of Cambridge, Madingley Rise, Cambridge, CB3 0HA, United Kingdom}

\author{Matthew Colless}
\affiliation{Research School of Astronomy and Astrophysics, Australian National University, Canberra, ACT 2611, Australia}

\author{Sarah Casura}
\affiliation{Hamburger Sternwarte, Universit\"{a}t Hamburg, Gojenbergsweg 112, D-21029 Hamburg, Germany}

\doi{10.1017/pasa.2020.32}

\received {dd Mmm YYYY}
\revised  {dd Mmm YYYY}
\accepted {dd Mmm YYYY}
\published{dd Mmm YYYY}


\begin{document}

%------------------------------------------------------------------------------%

% Abstract of the paper 250 words limit - 242
\begin{abstract}

We use the MaNGA integral-field spectroscopic survey of low-redshift galaxies to compare the stellar populations of the bulge and disc components, identified from their \Sersic\ profiles, for various samples of galaxies. Bulge dominated regions tend to be more metal-rich and have slightly older stellar ages than their associated disc dominated regions. The metallicity difference is consistent with the deeper gravitational potential in bulges relative to discs, which allows bulges to retain more of the metals produced by stars. The age difference is due to star formation persisting longer in discs relative to bulges.  Relative to galaxies with lower stellar masses, galaxies with higher stellar masses tend to have bulge dominated regions that are more metal-rich and older (in light-weighted measurements) than their disc dominated regions. This suggests high-mass galaxies quench from the inside out, while lower-mass galaxies quench across the whole galaxy simultaneously. Early-type galaxies tend to have bulge dominated regions the same age as their disc dominated regions, while late-type galaxies tend to have disc dominated regions significantly younger than their bulge dominated regions.  Central galaxies tend to have a greater metallicity difference between their bulge dominated regions and disc dominated regions than satellite galaxies at similar stellar mass. This difference may be explained by central galaxies being subject to mergers or extended gas accretion bringing new, lower-metallicity gas to the disc, thereby reducing the average metallicity and age of the stars;  quenching of satellite discs may also play a role.

\end{abstract}

%------------------------------------------------------------------------------%

\section{Introduction}

Galaxies naturally split into two principal stellar components: a bulge and a disc. In the Milky Way, the bulge is old ($>$10~Gyr) and metal-rich ([Z/H] \around 0.25) \citep{ness13}, while the disc has, on average, solar [Z/H] and a light-weighted age of a few Gyr \citep{casagrande11}; for more details of the properties of the Milky Way, see \citet{venn04}, \citet{blandhawthorn16}, \citet{duong19} and \citet{rojas-arriagada19}. Are these differences between the Milky Way bulge and disc typical of other galaxies? What are the mechanisms that produce these differences?

While initial studies of the stellar population of galaxies focussed on early-type galaxies, it has now been shown that it is possible to extend single stellar population fitting to galaxies with significant star formation \citep{peletier07,ganda07,sanchezblazquez14}. These studies found that the correlations observed in early-type galaxies, like the metallicity-mass relation, generally extend to late-type galaxies, though the scatter in the late-type population is significantly larger than in early-type galaxies.

Models of galaxy formation involving the monolithic collapse of a gas cloud usually produce relatively large [Z/H] gradients, decreasing from the centre of the galaxy \citep{matteucci89,goetz92,pipino10}; \citet{pipino10} typically find steep gradients of around $-$0.3\,dex in metallicity per decade variation in radius. In contrast, models involving multiple mergers produce much flatter gradient profiles, with slopes a factor of 2--3 smaller \citep{dimatteo09,taylor17}. 

There is an extensive literature supporting the conclusion that the inner regions of galaxies are more metal rich than the outer regions. Studies finding negative metallicity gradients in galaxies include \citet{moorthy06}, \citet{macarthur09}, \citet{sanchezblazquez11}, \citet{sanchezblazquez14},\citet{delgado15}, \citet{goddard17a}, \citet{zheng17}, \citet{santucci20} and \citet{zibetti20}. Studies reporting specifically that bulges are more metal-rich than discs include \citet{johnston12}, \citet{johnston14}, \citet{ganda07}, \citet{frasermckelvie18}, \citet{tabor19}, \citet{barsanti20} and \citet{johnston22}.  Nonetheless, the literature is not unanimous in finding more metals in the central regions of galaxies, with different results reported by  \cite{goddard17a}, \cite{ferreras19}, \cite{dominguezsanchez20} and \citet{pak21}. 

There is a similarly extensive literature showing that the stars in the inner regions of galaxies tend to be older. Studies finding negative age gradients in galaxies include \citet{sanchezblazquez14}, \citet{delgado15}, \citet{goddard17a}, \cite{zheng17} and \citet{dominguezsanchez20}. Studies reporting specifically that bulges are older than discs include \citet{frasermckelvie18}, \citet{pak21} and \citet{johnston22}. However, a substantial number of studies find that the bulge and disc have similar stellar ages \citep{tabor19,ferreras19,sanchezblazquez11,sanchezblazquez14,goddard17a,zibetti20,barsanti20,johnston22}, while others report the inner regions of galaxies are younger than the outer regions \citep{johnston12,johnston14,goddard17a,frasermckelvie18,santucci20}.

Thus, although most results support bulges being older and more metal-rich than discs, there are diverging results (the Appendix gives a more detailed summary of the conclusions reached by the various studies mentioned above). At least some of the variation in the results reported in the literature is likely to be due to different observational methods or measurement techniques: long slit versus integral field spectroscopy; different ways of measuring ages and metallicities; different ways of measuring gradients; also different definitions of bulges and discs.  Moreover, the differing galaxy sample properties, such as the distribution of stellar mass, morphology, or environment, also play a significant role in explaining the differing conclusions. 

The goal of this work is to create a comprehensive study of the stellar population properties of galaxy bulges and discs using a large sample of galaxies.  To this end, we take galaxy samples from the IFU SDSS-IV MaNGA survey spanning a wide variety of galaxy types and study the stellar properties of their bulges and discs. Rather than consider gradients, we compare instead the stellar ages and metallicities integrated over the bulge and disc components, determined directly from their \Sersic\ profiles. This approach has the significant advantage of always comparing the integrated stellar populations from consistently-defined morphological components of galaxies, rather than trying to map 'gradients' that may simply reflect the mixture of these components as a function of radius. The disadvantage is that it is harder to compare to the existing literature, which usually reports stellar population variations in terms of radial gradients.  While the 1D photometric profile decomposition used here does not split the bulge and disc light as precisely as a full 2D photometric image bulge-disk decomposition, it has the advantage (at present) of being applicable to larger, more representative samples of galaxies from which firm conclusions can be drawn. 

This paper has the following structure. In Section~\ref{data} the galaxy data is introduced, starting with a summary of the SDSS-IV MaNGA data (\S\ref{manga}), then detailing how the bulge and disc regions were selected (\S\ref{selecting_regions}), describing how the ages and metallicities were derived from the spectra using stellar population models (\S\ref{stellar_population_model}), the selection cuts of the sample (\S\ref{The_Selection_Cuts_of_the_Sample}) and providing the other properties of the sample (\S\ref{properties_of_the_sample}). Section~\ref{results} compares the stellar populations of bulges and discs, first across the entire sample (\S\ref{entire_sample}) and then for subsamples based on stellar mass, specific star formation rate, morphology, the \Sersic\ index of the bulge component, and environment (\S\ref{subsample}). Section~\ref{discussion} compares the results to the literature and discusses likely physical causes. Section~\ref{conclusion} summarises the conclusions of the paper. The Appendix contains a more detailed review of the literature studies referenced in the introduction. We adopt a value of 70\,km\,s$^{-1}$\,Mpc$^{-1}$ for the Hubble constant.

%------------------------------------------------------------------------------%

\section{Data}
\label{data}

%------------------------------------------------------------------------------%

\subsection{MaNGA Data}
\label{manga}

The Sloan Digital Sky Survey (SDSS) Mapping Nearby Galaxies at APO (MaNGA) is a large optical integral-field spectroscopy survey of low-redshift galaxies spanning a broad range in stellar mass, star formation rate, \Sersic\ index, and morphology \citep{bundy15, drory15}. It is being carried out using the 2.5m telescope at the Apache Point Observatory \citep[APO;][]{gunn06} using the BOSS spectrographs \citep{smee13}. There are 17 IFUs, which are deployed simultaneously across the 7\,deg$^2$ field of view. The IFUs range in diameter from 12~arcsec to 32~arcsec. The survey sample has a roughly flat logarithmic mass distribution of galaxies with stellar masses greater than 10$^9$\,\Msun\ \citep{law15,wake17}. The wavelength coverage is from 3500\,\AA\ to 10000\,\AA\ with spectral resolving power $R$\around 2000 giving an instrumental resolution of \around 60\kms. The spatial resolution is 2.5\,arcsec FWHM after combining the dithered images. The galaxies lie within the redshift range $0.01<z<0.15$ \citep{yan16b}. Observations are 3\,hours in length and the signal-to-noise ratio (SNR) is 4--8\,\AA$^{-1}$ at 1.5$R_{\rm e}$. Details on the data reduction pipeline can be found in \citet{law16} and \citet{yan16a}. For this analysis SDSS Data Release 16 (DR16) was used \citep{ahumada20}. Morphologies were available for most of the sample of galaxies from \citet{fischer19} using deep learning analysis. Elliptical galaxies were removed from the sample by removing galaxies that had TTYPE$<$0 and S0 probability P\_S0$<$0.5. Stellar masses were taken from the mangatarget table. These were the NSA Elepetro masses that are K-corrected fits from the elliptical Petrosian fluxes \citep{blanton11}. Star formation rates for the sample were obtained from SED fitting given in the GALEX-SDSS-WISE Legacy Catalog \citep[GSWLC,][]{salim16}. Environmental data for the galaxies was taken from the GEMA-VAC (Galaxy Environment for MaNGA Value Added Catalog).  This catalog also details whether the galaxies are centrals or satellites.  A catalog with the spectroscopic redshifts for each spaxel of the MaNGA data cubes \citep{bolton12,talbot18} was also used to correct the spectra for kinematic effects.

%------------------------------------------------------------------------------%

\subsection{Selecting Bulge and Disc Regions}
\label{selecting_regions}

%------------------------------------------------------------------------------%

\begin{figure}
  \includegraphics[width=\columnwidth]{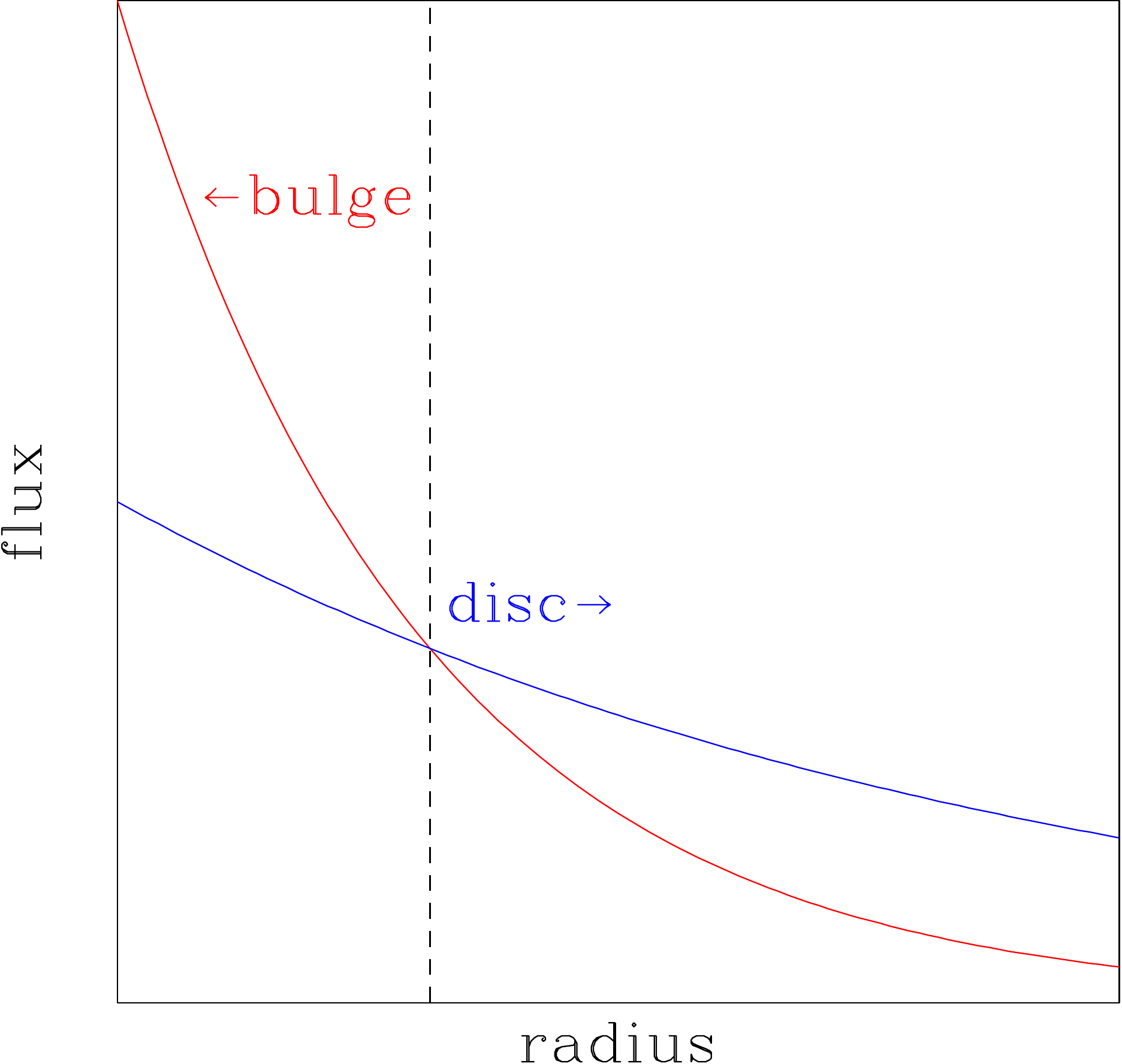}
  \caption{Definition of `bulge' and `disc' regions. The bulge is defined to be the bulge-dominated region, where the \Sersic\ luminosity profile of the bulge component is brighter than the \Sersic\ luminosity profile of the disc component; conversely for the definition of the disc (i.e.\ the disc-dominated region).  The figure is linear in flux.  Although the figure illustrates this in terms of the 1-dimensional profile, the definitions are actually based on the 2-dimensional profile.}
  \label{fig:plot_shape_new}
\end{figure}

%------------------------------------------------------------------------------%

Both one- and two-component \Sersic\ luminosity profile fits were available for this sample from the MaNGA PyMorph DR15 photometric catalogue \citep{fischer19} in g, r and i band. Usually the two-component \Sersic\ profiles have one profile with \Sersic\ index $n>1$ (the bulge) and the other $n=1$ (an exponential profile, the disc). However, some galaxies that have light profiles turning sharply downwards at large radii were fit with a $n=1$ profile for the central region and $n<1$ profile for the outer region. 

To select the regions defined as bulge and disc, the r band two-component model profiles were used. First, the point spread function for the observation was taken from the header parameter RFWHM for the MaNGA galaxies. This is the reconstructed FWHM in r-band and typically has value \around 2.5 arcsec.  The reconstructed FWHM in g-band and i-band usually differs from this value only in the 2nd decimal place.  Images were created based on the \Sersic\ profiles, with the bulge and disc \Sersic\ profile convolved with a Gaussian to take into account the smoothing due to the point spread function. The bulge-dominated (disc-dominated) regions are simply defined as those where the \Sersic\ profile representing the bulge is brighter (fainter) than the \Sersic\ profile representing the disc. Figure~\ref{fig:plot_shape_new} provides a visual representation of this as a function of radius, although the actual decomposition is done in two dimensions. Near the change-over between bulge- and disc-dominated regions, there is still significant light from the other component; this contamination blurs the results of our simple dichotomy somewhat and means that the true differences between the bulge and disc components are likely stronger than those measured here. The practical advantage of this method of decomposition is that it can be done at lower signal-to-noise than other methods and requires minimal assumptions about the properties of the disc and bulge. The interpretational advantage is that, whereas gradients conflate the admixture of the two components' stellar populations with their relative physical scales, this method compares the integrated stellar populations of the two components independently of their physical sizes.

Galaxies were removed from our sample if the bulge \Sersic\ profile was everywhere greater than the disc \Sersic\ profile, or vice versa. Galaxies were also removed if the disc \Sersic\ profile was greater than the bulge at the centre of the galaxy. Where there were repeat observations of the same galaxy, the data cube with the highest signal-to-noise in the disc spectrum was used.

The selected bulge and disc regions of the MaNGA data cubes are summed to convert them to one-dimensional spectra, increasing the signal-to-noise. One-dimensional spectra are also made from the summed spectra of the entire data cubes for comparison. 

%------------------------------------------------------------------------------%

\subsection{Stellar Population Model Fitting}
\label{stellar_population_model}

%------------------------------------------------------------------------------%

Stellar population ages and metallicities are measured from the summed, one-dimensional bulge and disc spectra using a full spectral fit based on theoretical stellar population models from the Medium resolution INT Library of Empirical Spectra \citep[MILES;][]{sanchez06,vazdekis10,vazdekis15}, BaSTI isochrones \citep{pietrinferni04, pietrinferni06}, and a \citet{chabrier03} initial mass function. Measurements of the stellar population ages and metallicities are also made on the combined overall galaxy spectrum. 

The process begins with de-redshifting the complete MaNGA spectrum. A first fit to the spectrum is then made using the penalised pixel-fitting code \citep[pPXF;][]{cappellari17} and the nominal variances in the spectrum. Based on the $\chi^2_{\rm reduced}$ of the first fit, the variances are rescaled to give $\chi^2_{\rm reduced} = 1$. With this improved noise estimate, a second pPXF fit is performed to identify any remaining bad pixels (including emission-line features) using 3~$\sigma$-clipping. The pixels identified as bad or containing emission lines are then replaced by the best-fit model from this second fit. The strong sky line at 5577\,\AA\ is masked (over 5565--5590\,\AA) and 13 common emission lines are also masked using the pPXF CLEAN keyword. 

After these pre-processing stages, the masked spectrum is fit using a linear combination of the single-instantaneous-burst MILES synthetic population templates and a degree-10 Legendre multiplicative polynomial. The role of the multiplicative polynomial is to correct the shape of the continuum and account for dust extinction. The MILES templates cover an approximately regular grid that spans a metallicity range $-2.27 \le$~[Z/H]~$\le 0.40$ and a log age range of -1.52~Gyr to 1.151~Gyr (unlogged age range of 0.03\,Gyr~$\le$~age~$\le 14.0$\,Gyr). The synthetic population templates have solar [$\alpha$/Fe] abundances. When fitting, each template is assigned a weight and the effective stellar age and [Z/H] are inferred from the combination of weights; both light-weighted and mass-weighted fits are performed. Light-weighted measurements are biased towards the light from bright, massive stars that have short lifespans; mass-weighted measurements are biased towards the stars that make up most of the stellar mass, which tend to have lower stellar masses and are long-lived. For more details on the process of stellar population fitting see \citet{barone20}, whose procedures we follow here.

The random errors on the ages and metallicities were generated by taking template spectrum and adding noise to them.  The stellar ages and metallicities were then measured from these noisy spectra and the variation in the results computed.  A relationship between error and signal-to-noise ratio  of the spectrum was created from this procedure and used to estimate the error for the ages and metallicities for the galaxy spectra.  Based on this work a minimum continuum signal-to-noise ratio of 15 per pixel was required in the bulge and disc spectra for the galaxies to be included in the stellar population sample.  At a signal-to-noise ratio of 15 per pixel, the error for the light-weighted parameters were 0.048\,dex in [Z/H] and 0.048\,log~Gyr in log age, while for the mass-weighted parameters the error were 0.056\,dex in [Z/H] and 0.055\,log~Gyr in log age.  Besides the random errors, there are systematic errors that affect the results that cannot be easily estimated.

%------------------------------------------------------------------------------%

\subsection{The Selection Cuts of the Sample}
\label{The_Selection_Cuts_of_the_Sample}

A total of 4857 data cubes were downloaded from DR16. Of these, 4672 have \Sersic\ profiles listed in the MaNGA pymorph tables. After removing galaxies that have undetermined \Sersic\ values for the bulge ($n = -999$), 4266 data cubes remain. Galaxies that have \Sersic\ profiles offset more than 2.5~arcsec from the centre of the data cubes were removed, leaving 4239 data cubes.  Of these 712 had their bulge \Sersic\ profile always larger than the disc.  531 of these are elliptical galaxies and the rest of them should probably be labelled as elliptical galaxies as they have a not very significant disc.  These were removed from the sample leaving 3527.  1137 galaxies had their disc \Sersic\ profile always larger than the bulge and 229 had the disc \Sersic\ profile larger than the bulge at the centre of the galaxy.  These galaxies are probably best labelled as galaxies without bulges.  These 1366 cases were removed from the sample, leaving 2161 data cubes. For some data cubes the stellar population analysis failed due to poor signal-to-noise in the spectra, giving 2026 useable data cubes. At this point, multiple observations of the same galaxy were removed, by adopting the data cube with the highest signal-to-noise in the disc spectrum, leaving 2007 unique galaxies. A signal-to-noise cut in both the bulge and disc spectra of 15 per pixel reduced the sample of galaxies to 1361. Selecting galaxies that had a measurable ESWEIGHT for completeness corrections left 1342 galaxies. Removing galaxies classified as ellipticals from the sample reduced this to 1164. This sample was visually inspected for multiple objects in the MaNGA field of view. There were 106 galaxies with a companion within the field of view, either a star, a nearby galaxy, or a hot spot in the galaxy.  As these could potentially bias the measurement of the stellar population of the disc, these objects were removed from the sample, leaving 1058 galaxies. Star formation rates are available for 891 of these galaxies, nearest neighbour measurements for 762, and central/satellite classifications for 1022.

%------------------------------------------------------------------------------%

\subsection{Other Properties of the Sample}
\label{properties_of_the_sample}

%------------------------------------------------------------------------------%

\begin{figure}
  \includegraphics[width=\columnwidth]{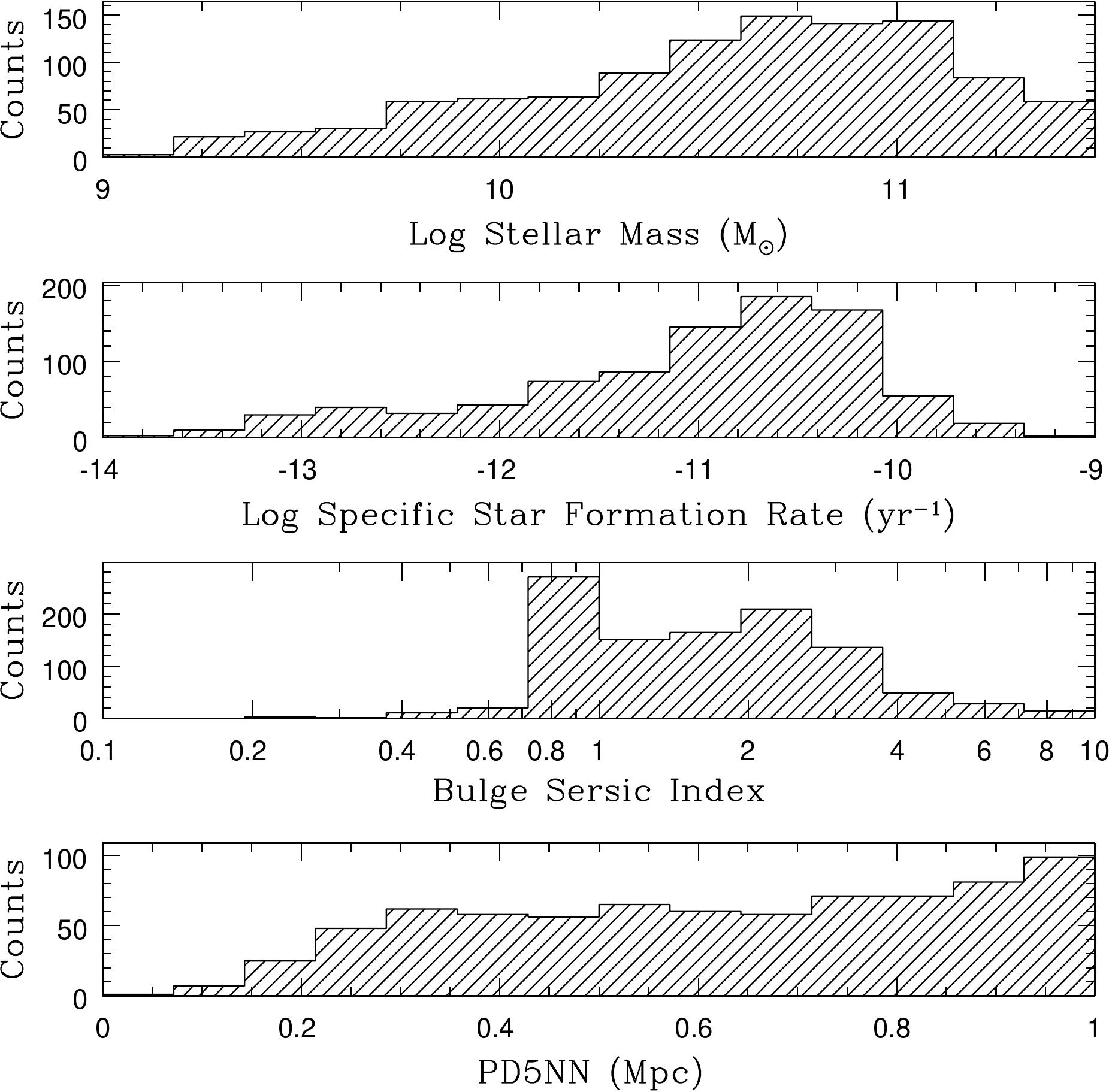}

  \caption{The unweighted histograms of various properties of the galaxies in the final sample. Top to bottom: (i)~log stellar mass; (ii)~log specific star formation rate; (iii)~\Sersic\ indices for the galaxy bulges; and (iv)~projected distance to the 5th nearest neighbour (PD5NN) in Mpc, as a measure of environmental density.}

  \label{fig:hist_ssp_bulge_disk_combined}
\end{figure}

%------------------------------------------------------------------------------%

Figure~\ref{fig:hist_ssp_bulge_disk_combined} shows the distributions of various properties for galaxies in the final sample. The top panel shows that the sample has a peak in log stellar mass at $\log\,M/M_\odot \sim 10.85$, with a long tail to lower masses. The second panel shows the specific star formation rate (sSFR) distribution has a shape similar to the log stellar mass distribution, with a peak at $\log({\rm sSFR/year}) \sim -10.75$ and a long tail to lower specific star formation rates. The third panel gives the distribution of the \Sersic\ index of the bulge, showing many galaxies with `bulge' \Sersic\ index of 1, the fixed index used for galaxies with light profiles that go sharply downwards at large radii (as mentioned previously). The bottom panel shows the projected distance to the 5th nearest neighbour (PD5NN) in Mpc, a measure of the local environmental density; there are few galaxies at the highest densities (PD5NN below 0.2 Mpc), indicative of clusters, but more in lower-density regions. 

\begin{figure}
  \includegraphics[width=\columnwidth]{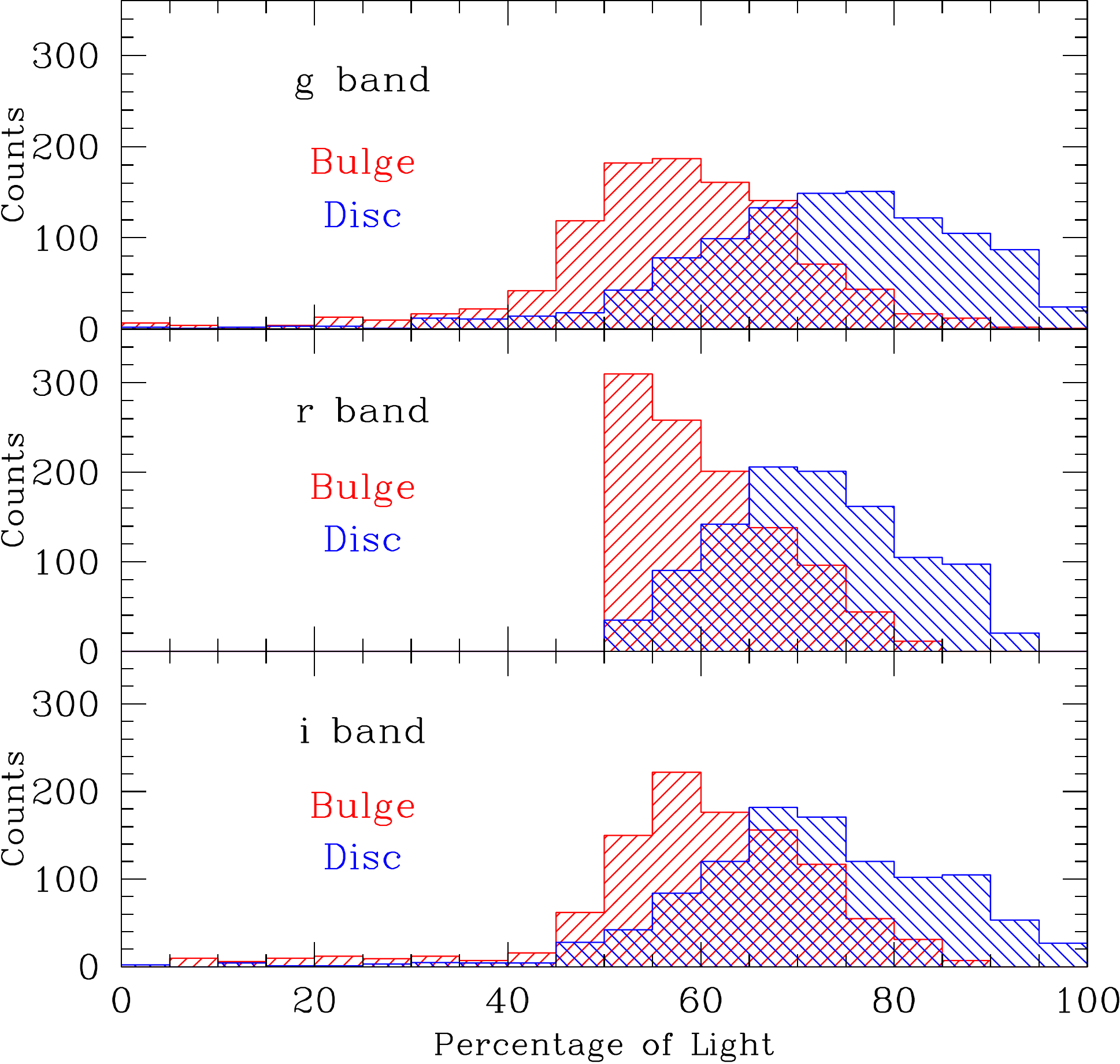}
  \caption{The number of galaxies with a given percentage of light contributed by the bulge in the bulge-dominated region is shown by the red histogram in g, r and i band. The number of galaxies with a given percentage of light contributed by the disc in the disc-dominated region is shown in the blue histogram in g, r and i band. The remaining light in each region is the contaminant from the other component, assuming that the \Sersic\ profile accurately traces the light from the bulge and disc. The median contribution from the bulge in the bulge-dominated region in r band is 59\%, while the median contribution from the disc in the disc-dominated region in r band is 71\%}.
  \label{fig:hist_ssp_bulge_disk_Sersic_percentage}

\end{figure}

The red histograms in Figure~\ref{fig:hist_ssp_bulge_disk_Sersic_percentage} show the distribution of the fraction of light in the bulge region (i.e.\ where the bulge profile dominates) that actually belongs to the bulge component (as computed from the two-component fit) for g, r and i band. The blue histogram in the figure shows the distribution of the fraction of light in the disc region (i.e.\ where the disc profile dominates) that actually belongs to the disc component for the same bands.  By definition there is nothing below 50\% in the r band sample as these galaxies would have been classified as bulge or disc dominated and removed from our sample.  As the selection is being done in r band it is possible for the g and i band measurements to be below 50\%.  The very low percentages for some of the g and i band galaxies are when the \Sersic\ profiles differ substantially between bands, usually in the orientation of the galaxy. 

The level of cross-contamination between components is significant (it should be noted that this is the amount of light in each region according to the \Sersic\ profiles; the actual stellar population of the bulge and disc may not follow the profiles exactly.) This contamination of bulge by disc and disc by bulge will tend to blur out any physical differences between bulge and disc properties, so the actual trends can only be stronger.

There is a wavelength-dependant component to contamination, with the bulge having higher percentages for i band over r band and with the disc having higher percentages for g band over i band.  This is similar to the differences in the bulge to total ratio found by \citet{schulz03} for U, B, V, I bands, i.e. increasing bulge to total ratio as one moves to redder bands.  This wavelength contamination shift will effect the stellar population parameters measured especially as they refer to other properties like \Sersic\ index, stellar masses and specific star formation that also vary with wavelength.  To quantify this effect stellar population models were run on bulge and disc spectra selected using the g and i band \Sersic\ profiles rather than r band.  This is discussed in Section~\ref{subsample} after the main results have been presented.

%------------------------------------------------------------------------------%

%------------------------------------------------------------------------------%

%------------------------------------------------------------------------------%

\section{Comparison of bulge and disc stellar populations}
\label{results}
%------------------------------------------------------------------------------%

\subsection{Results for the entire sample}
\label{entire_sample}

%------------------------------------------------------------------------------%

\begin{figure*}
  \includegraphics[width=8.5cm]{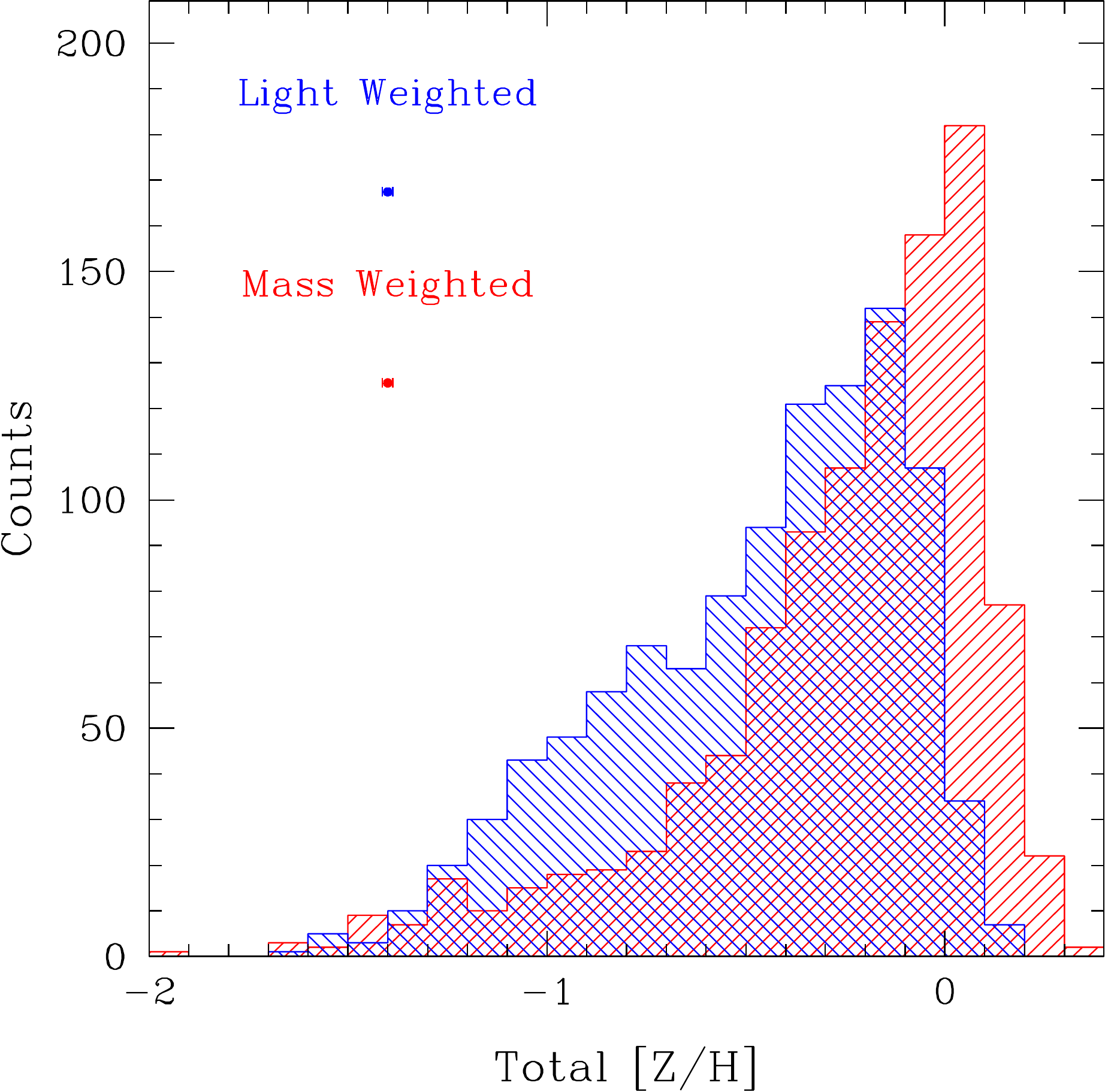} 
  \hspace*{0.5cm}
  \includegraphics[width=8.5cm]{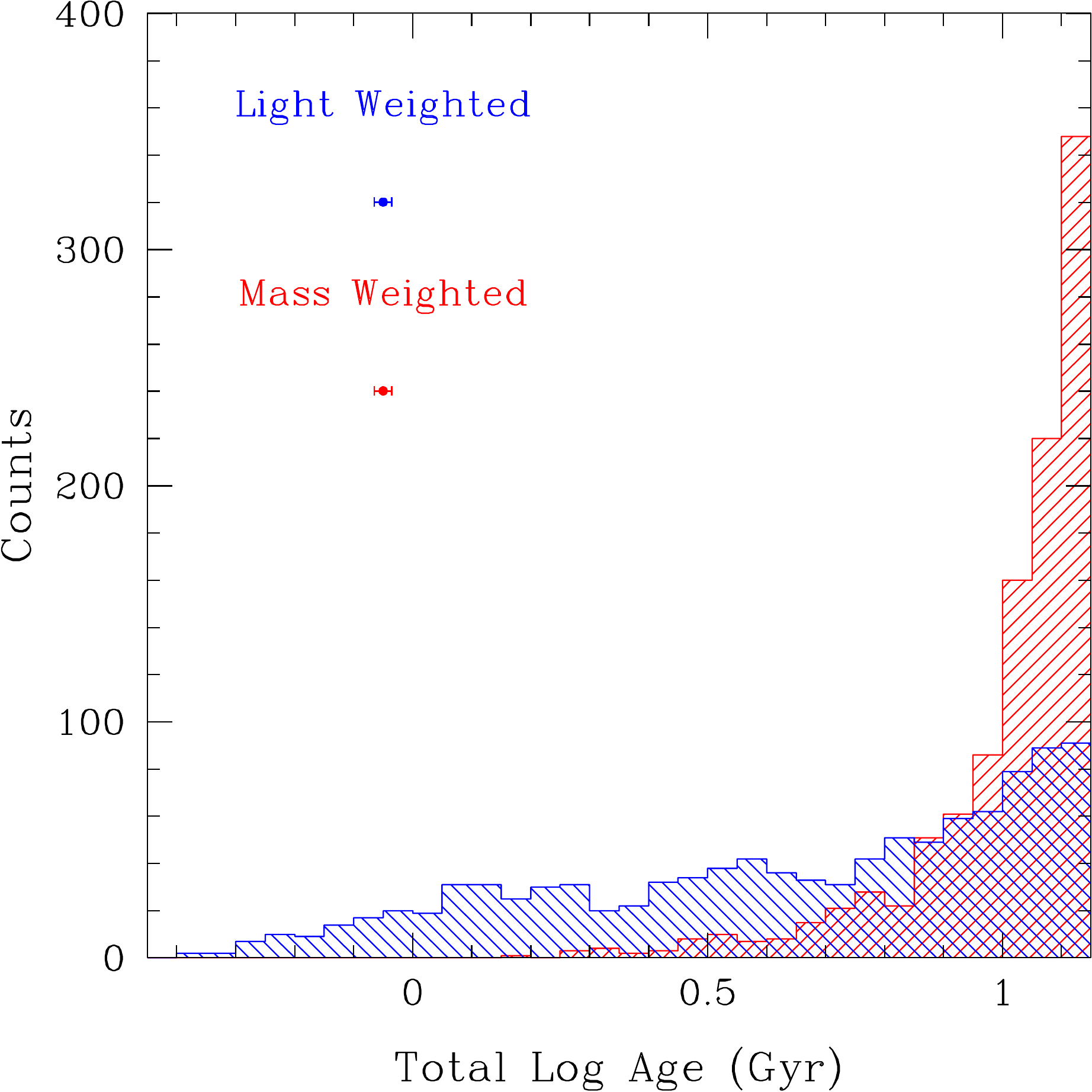}

  \caption{Left: Histogram of overall stellar [Z/H] for galaxies in the sample. The blue histogram is the overall light-weighted stellar [Z/H] and the red histogram is the overall mass-weighted stellar [Z/H]. Right: Histogram of overall stellar log ages for galaxies in the sample. The blue histogram is the overall light-weighted stellar log ages and the red histogram is the overall mass-weighted stellar log ages.  In both panels the median errors for the sample are displayed.  This is the random error.  There is a larger systematic error that is not as easily calculated.  The median signal to noise for the sample is 52.}
  \label{fig:hist_ssp_bulge_disk_Z}
  \label{fig:hist_ssp_bulge_disk_age}
\end{figure*}

The left panel of Figure~\ref{fig:hist_ssp_bulge_disk_Z} shows the distribution of the overall stellar [Z/H] for the sample of galaxies, with the blue histogram showing the light-weighted [Z/H] and the red histogram the mass-weighted [Z/H]. These overall stellar [Z/H] values are measured from the spectrum for the whole galaxy before the decomposition into bulge and disc regions. The light weighted distribution is offset from the mass weighted distribution and the mass weighted distribution has more values at higher metallicity than the light weighted distribution.  Both have a sharp decline at higher metallicities with long tails at lower metallicities.

%------------------------------------------------------------------------------%

%------------------------------------------------------------------------------%

The blue histogram in the right panel of Figure~\ref{fig:hist_ssp_bulge_disk_age} shows the distribution of the overall light-weighted stellar log age for the sample of galaxies from the spectra for the whole galaxy before decomposition into bulge and disc samples.  The sample has a fairly uniform distribution until you reach the highest ages where there is a definite increase. The red histogram in the right panel of Figure~\ref{fig:hist_ssp_bulge_disk_age} shows the distribution of the overall mass-weighted stellar age for the sample of galaxies; again, this is from the spectra for the whole galaxy before decomposition into bulge and disc samples. This distribution is strongly peaked at high log ages.  The differences between the light- and mass-weighted stellar metallicities are fairly modest but for stellar ages there are very different results for light-weighted and mass-weighted ages. The era of most recent star formation, as determined from the high-mass stars measured by the light-weighted ages, varies across the full range of stellar ages and indicates when the last major burst of star formation occurred, whereas the era of peak star formation, as determined from the low-mass stars measured by the mass-weighted ages, is strongly biased towards very old ages.

%------------------------------------------------------------------------------%

\begin{figure*}
  \includegraphics[width=8.5cm]{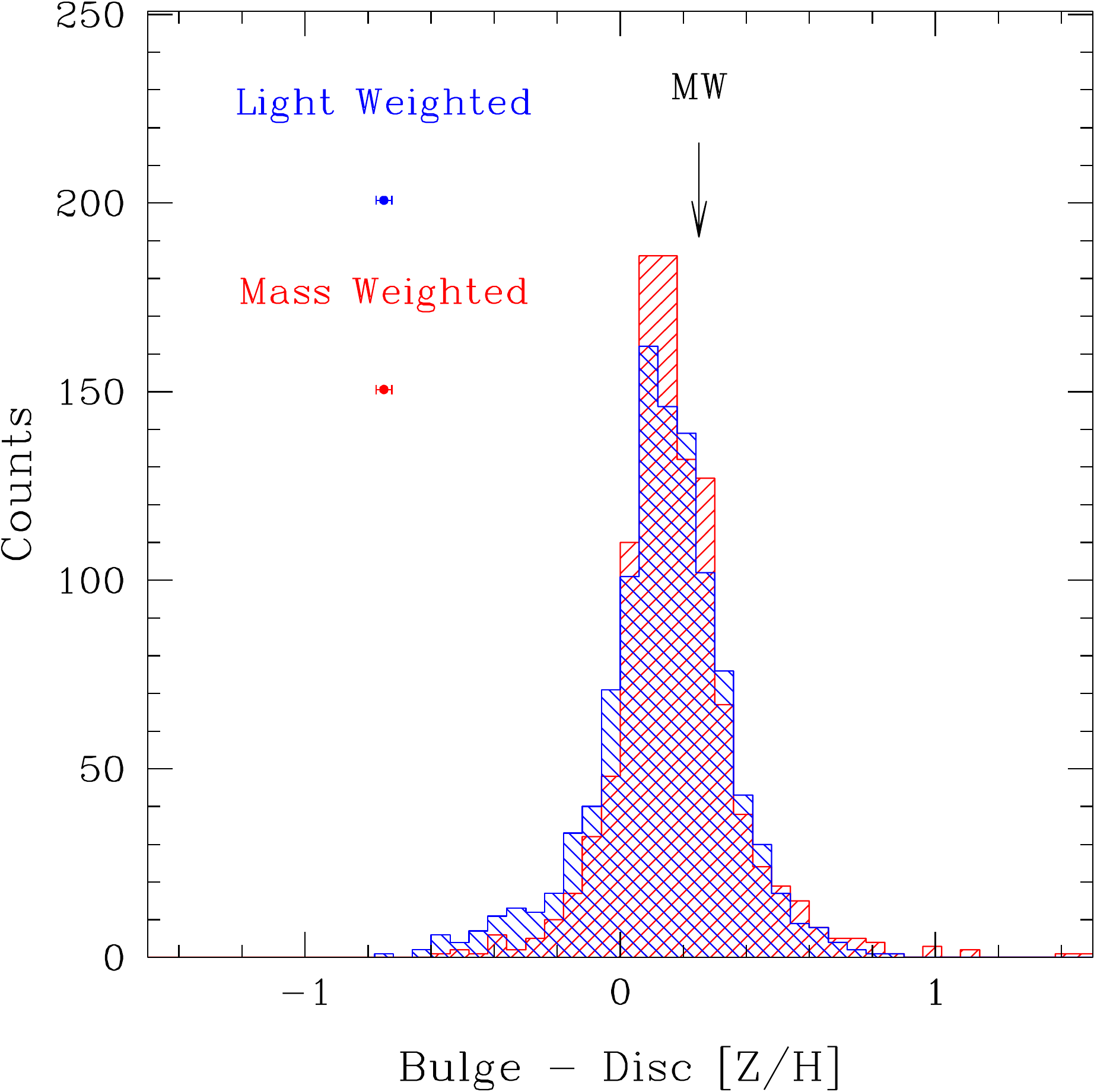}
  \hspace*{0.5cm}
  \includegraphics[width=8.5cm]{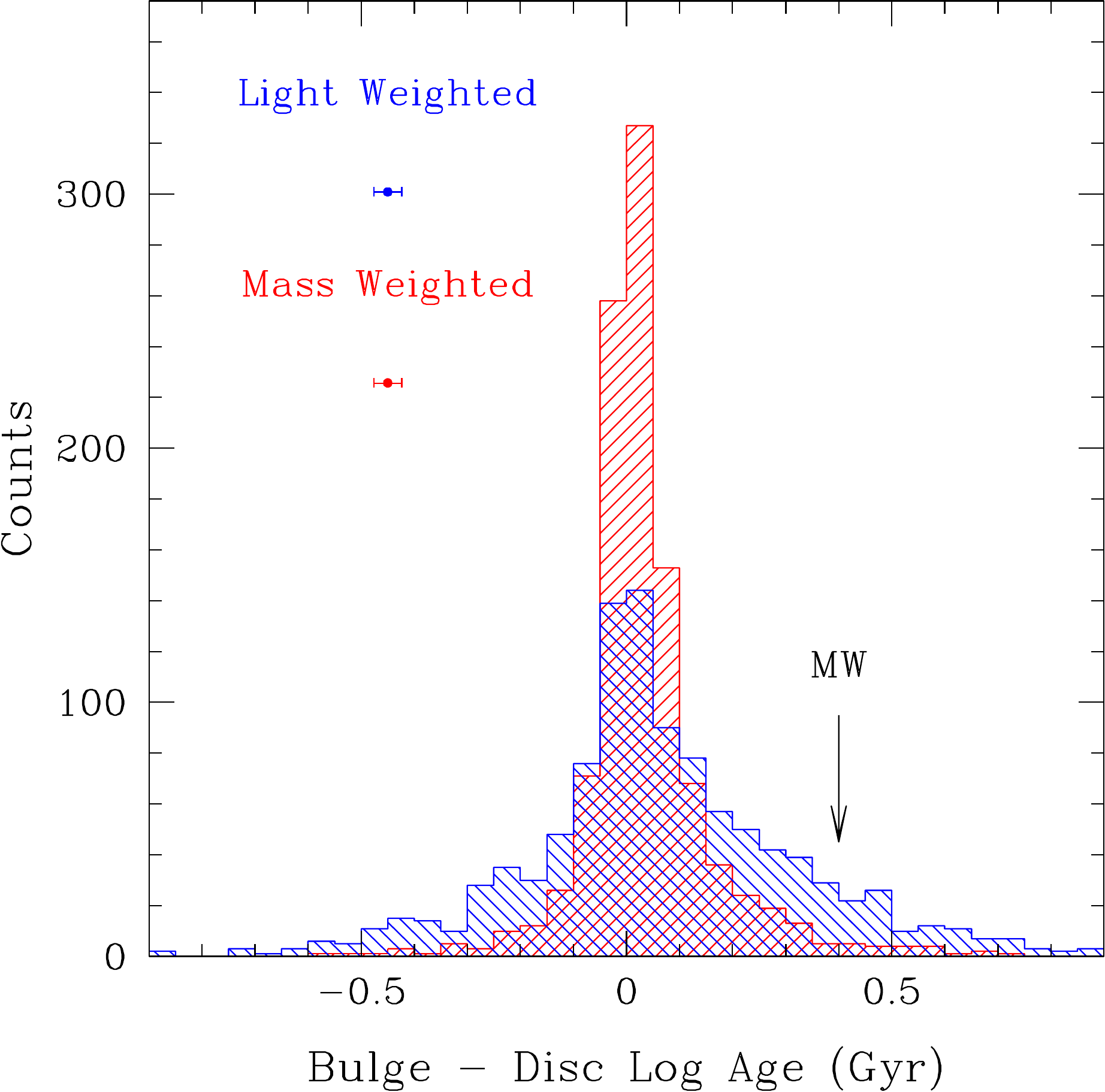}

  \caption{Left: Histogram of differences between the stellar [Z/H] of the bulge and the disc for the galaxies in the sample. The blue histogram shows the differences in light-weighted [Z/H]; the red histogram shows the  differences in mass-weighted [Z/H]. The arrow is the approximate value of the difference between bulge [Z/H] and the disc [Z/H] for the Milky Way.  The error bars are the median error bars for the sample.  Right: Histogram of the differences between the stellar log ages of the bulge and the disc for the galaxies in the sample. The blue histogram shows the differences between the light-weighted stellar log ages; the red histogram shows the differences between the mass-weighted stellar log ages. The arrow is the approximate value of the difference between bulge age and the disc age for the Milky Way \citep{casagrande11,ness13}.  The error bars are the median error bars for the sample.  The median signal-to-noise for the bulge sample is 62 and for the disc sample is 32.}

  \label{fig:hist_ssp_bulge_disk_diff_Z}
  \label{fig:hist_ssp_bulge_disk_diff_age}
\end{figure*}

%------------------------------------------------------------------------------%

The blue histogram in the left panel of Figure~\ref{fig:hist_ssp_bulge_disk_diff_Z} shows the distribution of differences between the bulge and disc light-weighted stellar metallicities for the sample of galaxies. The differences are generally positive and the distribution is strongly peaked around [Z/H]\,$\sim$\,0.2\,dex, showing that the bulge is usually slightly more metal-rich than the disc for this sample. The red histogram in the left panel of  Figure~\ref{fig:hist_ssp_bulge_disk_diff_Z} shows the distribution of the differences between the bulge and disc mass-weighted stellar metallicities for the sample of galaxies. The distribution is similar to the light-weighted one, with a peak around [Z/H]\,$\sim$\,0.2\,dex. As shown by the arrow in this figure, the Milky Way is a fairly typical galaxy in terms of the metallicity difference between its bulge and its disc as derived from \citet{casagrande11} and \citet{ness13}.

The blue histogram in the right panel of Figure~\ref{fig:hist_ssp_bulge_disk_diff_age} shows the distribution of the difference between the bulge and disc light-weighted stellar log ages for the sample of galaxies.  The peak of the distribution is slightly above zero with well defined wings at greater differences. The red histogram in the right panel of Figure~\ref{fig:hist_ssp_bulge_disk_diff_age} shows the distribution of the difference between the bulge and disc mass-weighted stellar log ages for the sample of galaxies. This distribution also peaks slightly above zero but the distribution is more peaked and narrower than the light-weighted distribution.  As shown by the arrow in this figure, the Milky Way, with its bulge significantly older than its disk, is a somewhat atypical galaxy in terms of the age difference between bulge and disc, lying in the positive tail of the distribution \citep{casagrande11,ness13}. It should be noted here that the value for the Milky Way is very uncertain and that different methods were used to measure the ages in the Milky Way and in the galaxy sample (with the Milky Way bulge and disc both showing a range in stellar ages, $\pm 2$~Gyr). However, the comparison is interesting, despite its limitations.

%------------------------------------------------------------------------------%

\subsection{Results for subsamples}
\label{subsample}

%------------------------------------------------------------------------------%

The number of galaxies in the sample is sufficiently large that it can be broken down into subsamples that still have reasonable statistical power. We split the full sample into contrasting pairs of subsamples based on: (i)~bulge \Sersic\ index, with a split at $n = 1.5$; (ii)~specific star formation rate, with a split at $10^{-11}$\,yr$^{-1}$; (iii)~stellar mass, with a split at $M_\star = 5 \times 10^{10} M_\odot$; (iv)~visual morphology, split into early types and late types; (v)~environment, as determined by the projected distance to the 5th nearest neighbour, with a split at PD5NN\,=\,0.7\,Mpc; and (vi)~whether the galaxy is a central or satellite galaxy. The \Sersic, stellar mass and projected distance splits were chosen so there would be roughly equal numbers of galaxies in each category. The sSFR split was chosen to try to separate the strongly star-forming main sequence from weakly star-forming/quiescent galaxies, while still having large enough samples for statistics. 

For each subsample, the weighted median values for the stellar population results for the sample of galaxies were determined, with weights correcting using the MaNGA esweights (i.e.\ the medians are for volume-limited samples). The errors on the weighted medians were estimated using a bootstrap method: a random sample of values the same size as the observed sample was selected from the observed values (with replacement) and the weighted median calculated for this selection; this was repeated 100 times and the standard deviation of these weighted median values was used as the estimate of the standard error on the weighted median.

%------------------------------------------------------------------------------%

\begin{figure*}
  \includegraphics[width=8cm]{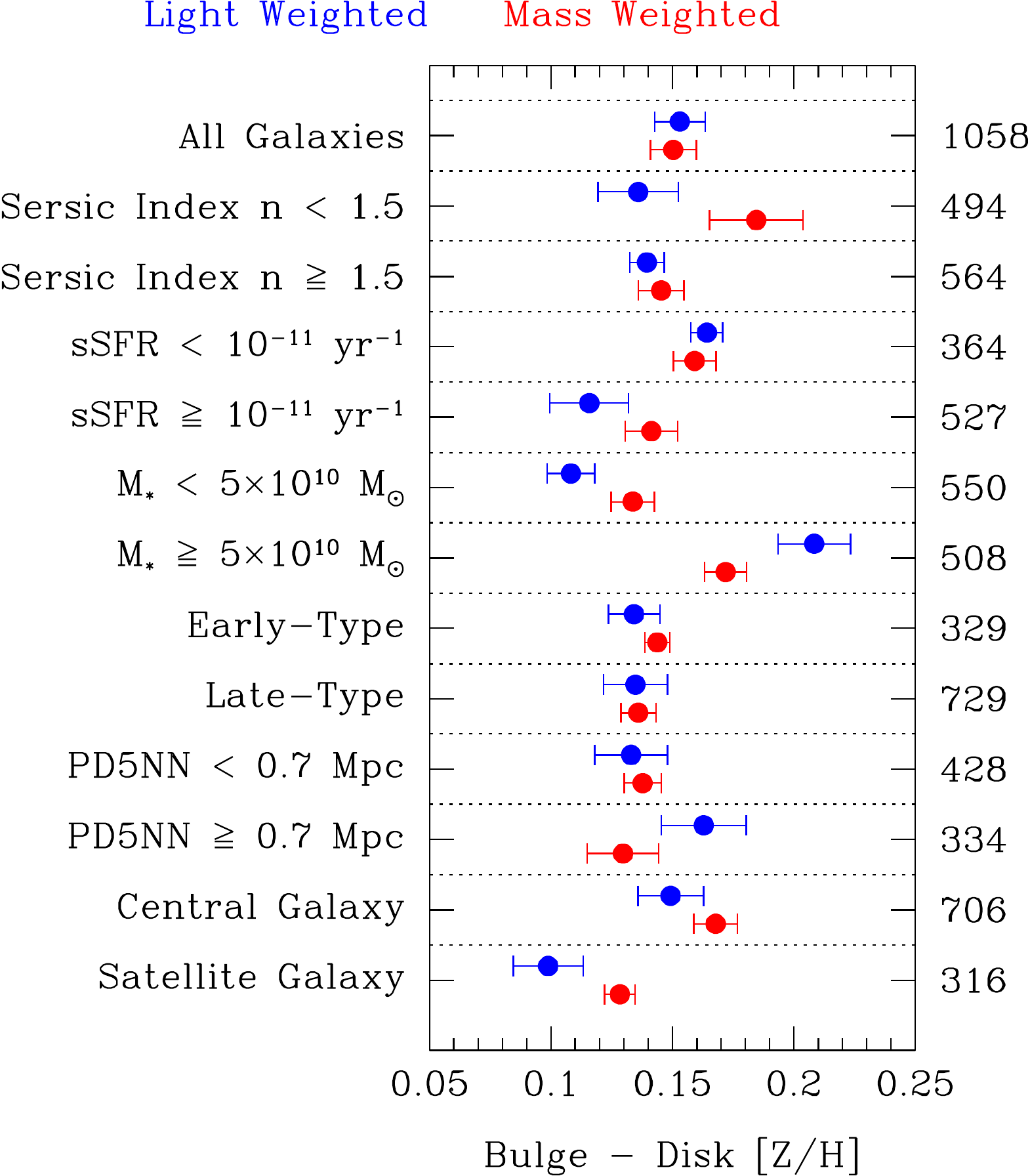}
  \hspace*{0.5cm}
  \includegraphics[width=8cm]{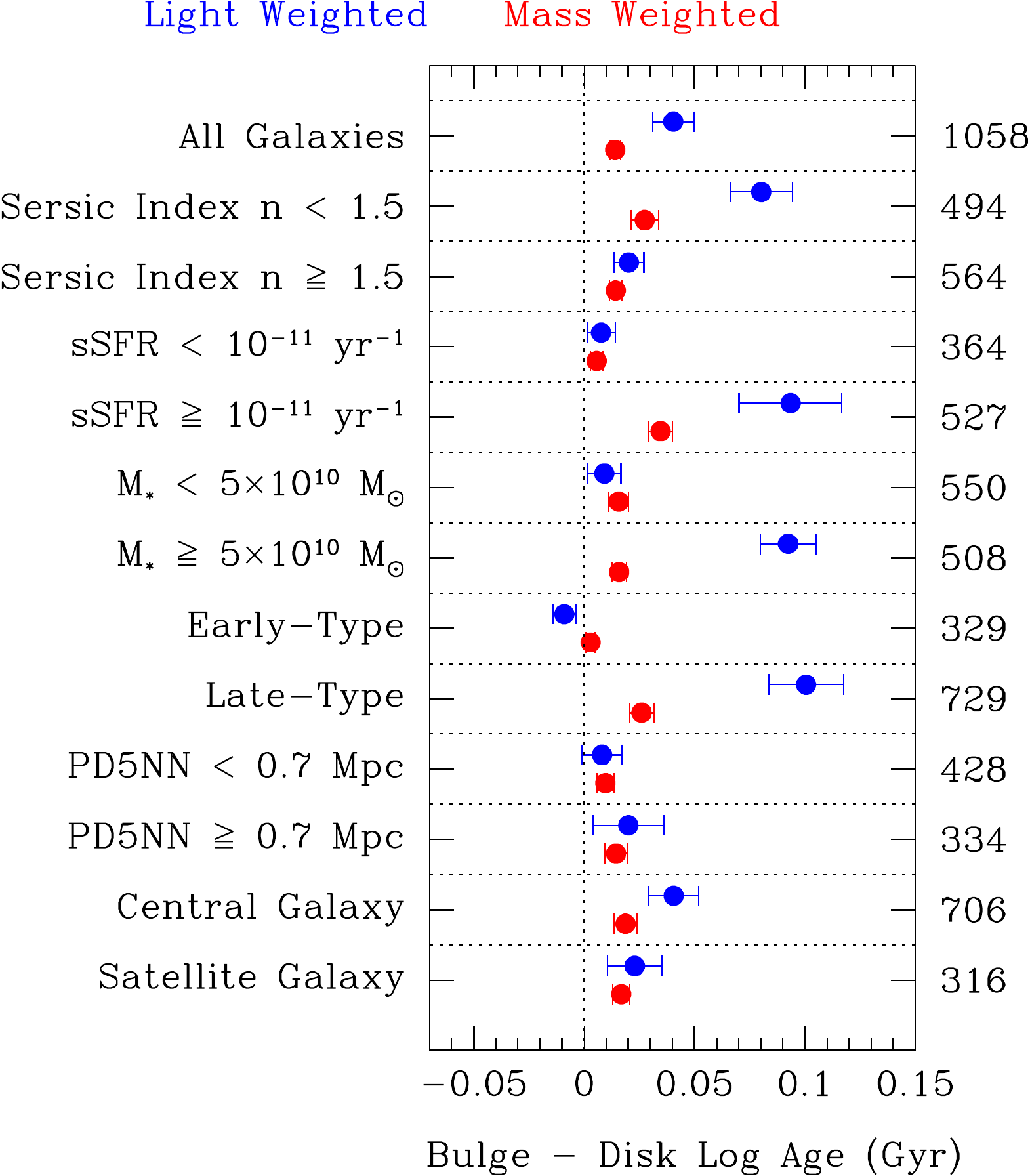}

  \caption{Left: The volume-corrected median values (and estimated uncertainties) for the difference in stellar metallicity between the bulge and disc for various subsamples. The blue points are the median values of the difference in the light-weighted stellar [Z/H]; the red points are the median values of the difference in the mass-weighted stellar [Z/H]. Right: The volume-corrected median values (and estimated uncertainties) for the difference in stellar log age between the bulge and disc for various subsamples. The blue points are the median values of the difference in the light-weighted stellar age; the red points are the median values of the difference in the mass-weighted stellar age. In all cases the number of galaxies in each sample is written on the right. PD5NN is the projected distance to the 5th nearest neighbour.}

  \label{fig:plot_measured_quantities_z_diff}
  \label{fig:plot_measured_quantities_age_diff}
\end{figure*}

%------------------------------------------------------------------------------%

The left panel of Figure~\ref{fig:plot_measured_quantities_z_diff} shows the volume-corrected median differences in the stellar metallicities, [Z/H], of the bulge and disc in the various subsamples of galaxies, for light-weighted (blue) and mass-weighted (red) estimates. For the whole sample, the bulges are more metal-rich than discs by around 0.15\,dex for both light-weighted and mass-weighted measurements. This pattern of the bulge having more metals than the disc continues throughout the subsamples.  Subsamples based on \Sersic\ index mostly agree well with the entire sample measurements, except for the mass-weighted measurement for galaxies with low $n$, where the difference is higher than the other measurements (by $1.9 \sigma$ larger).  The light-weighted high specific star formation sample has higher metallicity difference than the low specific star formation sample (by $2.5 \sigma$ larger). The mass-weighted specific star formation values agree with each other.  Both the light-weighted and mass-weighted metallicity measurements for high stellar mass samples are higher than the low stellar mass measurements  (light-weighted $7.3 \sigma$ higher, mass-weighted $4.2 \sigma$ higher).  There is no statistical difference between the early- and late-type subsamples, nor between subsamples based on the projected distance to the 5th nearest neighbour (PD5NN). There is a difference between the central galaxy and satellite subsamples, with centrals having more metal-rich bulges than discs relative to satellites (light-weighted $3.6 \sigma$ higher, mass-weighted $4.3 \sigma$ larger).  One might suspect this trend is due to central galaxies tending to be more massive than satellite galaxies, but the distributions of stellar masses for centrals and satellites in our sample are similar, so the difference seen here is not simply a mass effect.

The right panel of Figure~\ref{fig:plot_measured_quantities_age_diff} shows the volume-corrected median differences in the stellar ages of the bulge and disc in the various subsamples of galaxies, for light-weighted (blue) and mass-weighted (red) estimates. For all galaxies, for both light-weighted and mass-weighted measurements, the bulges are slightly older than the discs with the light-weighted measurement higher than the mass-weighted measurement (the light-weighted measurement is $3.9 \sigma$ above zero and the mass-weighted measurement is $6.5 \sigma$ above zero). For subsamples based on \Sersic\ index, most agree with each other except the low \Sersic\ light-weighted measurement which is larger than the others ($4.1 \sigma$ larger).  The higher specific star formation sample has larger age difference than the low specific star formation sample with the light-weighted sample having the greater difference  (light-weighted $3.9 \sigma$ larger, mass-weighted $4.7 \sigma$ larger).   While the mass-weighted sample shows no significant difference between the high and low stellar mass samples the light weighted sample does, with the high stellar mass sample having a larger log age difference than low stellar mass sample ($5.8 \sigma$ larger).  Early-type galaxies tend to have bulges of similar ages to their disc, while late-type galaxies have bulges significantly older than their discs (light-weighted $6.6 \sigma$ larger, mass-weighted $4.2 \sigma$ larger). There is no trend with projected distance to the 5th nearest neighbour, with both low- and high-density regions having similar medians. Centrals and sattelites have similar age differences between their bulges and discs.

It is interesting to return to the values for the Milky Way at this point, as we can now compare directly to the subsample that best describes our Galaxy. The Milky Way has a measured star formation rate of $1.65 \pm 0.19$\,\Msun\,yr$^{-1}$ and a stellar mass of $(6.08 \pm 1.14) \times 10^{10}$\,\Msun \citep{licquia15}. This gives it a specific star formation rate of $(2.71 \pm 0.60) \times 10^{-11}$~yr$^{-1}$. The Milky Way thus falls in our high-stellar-mass and high-specific-star-formation-rate subsamples. The [Z/H] difference of around 0.25\,dex \citep{ness13,casagrande11} between the bulge and disc of the Milky Way is slightly higher than, but still close to, the median values for these samples; this is also true of the late-type and central subsample measurements. In terms of [Z/H] difference between bulge and disc, the Milky Way galaxy is close to a representative example. By contrast, the age difference between the bulge and disc of the Milky Way, at around log age of 0.4\,Gyr \citep{ness13,casagrande11}, is significantly larger than the median value for the high sSFR, high stellar mass, late-type or central subsamples (for either light-weighted or mass-weighted measurements). Galaxies with bulge--disc age differences as high as the Milky Way are uncommon in our volume-corrected sample. \citet{delgado15} found that Milky Way-like galaxies stand out as those with the steepest radial age profiles. Additionally, the Milky Way has a rather uncommon merging history, with two fairly large current satellites and no mergers for the past several Gyr \citep{evans20}.

All of these results are affected by the contamination from the overlap of the bulge and disc stars, which will tend to reduce the apparent stellar population differences between the `bulge-dominated' and `disc-dominated' regions. It is therefore likely that the underlying physical differences are in fact larger than those we measure. The Milky Way's bulge--disc stellar age difference may not be as significant an outlier if this contamination effect is taken into account.

The selection here was done using the r band \Sersic\ profiles.  As mentioned above in Section~\ref{properties_of_the_sample} this analysis was repeated using selection with the g and i band \Sersic\ profiles.  The values for the medians were similar for the g and i band selections nearly always within $1 \sigma$ of the r band medians except for a few cases where they were within $2\sigma$.  There was no clear pattern in the values moving from g to r to i band.  The band used to select the spectra seems to have minimal effect on the stellar population results.

%------------------------------------------------------------------------------%

\section{Discussion}
\label{discussion}

The result for the full galaxy sample, that bulges have more metals than discs, agrees well with what is generally found in the literature \citep{johnston12,johnston14,tabor19,moorthy06,macarthur09,sanchezblazquez11,sanchezblazquez14,delgado15,goddard17a,zheng17,zibetti20,barsanti20,parikh21,johnston22}. Although most literature studies measure the gradient of metallicity with respect to radius, the sign of this gradient can broadly be interpreted in terms of comparing bulge-dominated and disc-dominated regions.   \citet{barone18,barone20} established that for galaxies as a whole the gravitational potential is the primary regulator of stellar metallicity, through its relationship to the gas escape velocity.  It is a small step to apply this result to the bulge (with its higher gravitational potential) and disc (with its lower gravitational potential) to explain the difference seen here in their metallicity. 

The higher value in bulge--disc metallicity difference for mass-weighted measurements of low-\Sersic-index galaxies is an artifact of the completeness correction.  While in general taking the medians without the completeness correction gives results similar to the ones with the correction, this result in the exception.  Without the completeness correction the bulge--disc metallicity difference for mass-weighted measurements of low-\Sersic-index galaxies is similar to the other results.  It is unclear why this anomaly occurs: there is no obvious physical explanation and the result is only $1.9 \sigma$, so it may just be random chance.

The light-weighted high specific star formation sample has a larger metallicity difference than the low specific star formation sample, though it is only a $2.5 \sigma$ larger.  This difference is not present in the mass-weighted measurements.  This may be an artifact of the sample, where our low sSFR sample tends to have higher stellar mass (i.e. most of the low stellar mass galaxies in our sample have high sSFR).  So this trend may just be a reflection of the pattern seen for stellar mass.

The higher metallicity difference between bulge and disk for high-stellar-mass galaxies is likely related to the difference in gravitational potential as discussed above.  The bulges of higher-mass galaxies are better able to hold onto their metals than lower-mass galaxies that have smaller bulges with lower gravitational potentials. The light-weighted measurement being higher than the mass-weighted one indicates that the high-mass stars formed from gas of higher metallicity that had been retained by the galaxy. These high-mass stars, having shorter lifetimes, likely formed after the majority of the low-mass stars that dominate the mass-weighted measurement and so were enriched by previous generations of stars. 

This trend with stellar mass appears to be contrary to the results of \citet{dominguezsanchez20}, who found that galaxies with masses above $\rm 3 \times 10^{10} M_\odot$ show little or no metallicity gradients, while less-massive galaxies in their sample tend to show significant metallicity gradients. Similarly, \citet{santucci20} found negative [Z/H] radial gradients in their sample, which show evidence of becoming shallower with increasing stellar mass.  The reason for this difference is unclear.  The significant difference between our work and theirs is their use of gradients versus our bulge-disc measurements.  Our bulges are allowed to vary in size, while the gradients ignore this variation.  However, above-mentioned results based on gradients have agreed with what we found using our method.  Our result does have the natural explanation of increasing gravitational potential with stellar mass as the source of the differences we see. 

We agree with \citet{pak21} in finding no trend of bulge--disc metallicity difference with environment, as measured by projected distance to the 5th nearest neighbour.  However, it should be noted that we are not probing the very dense environments that are seen in clusters, where there could be such an effect. 

There is a distinction between central and satellite galaxies, with centrals having higher metallicity difference between bulges and discs. In general central galaxies tend to be larger than satellite galaxies.  However in our sample this effect is not present, so cannot be the origin of the distinction.  Centrals also tend to have more mergers than satellites and accrete more gas.  This could be an explanation of what we see here with the mergers and the accretion bringing in fresh gas of lower metallicity that dilutes the metallicity of stars produced in the disc of centrals.  Another difference is that satellites undergo more quenching than centrals.  Star formation could have quenched in satellite discs before they could produce the same metallicity difference seen in centrals.  It is not possible to distinguish between these two possibilities with the data we have.

This result differs from that found by \citet{santucci20}, that metallicity gradients for fixed stellar mass were similar for satellites and central galaxies.  The most significant difference between our work and \citet{santucci20} is that they limit themselves to passive galaxies only, while we have star forming galaxies in our sample.  

We find in general that the bulges of galaxies are slightly older than their discs for both light-weighted and mass-weighted measurements. This is consistent with the measurements made by \citet{sanchezblazquez14,delgado15,goddard17a,zheng17,frasermckelvie18,pak21,johnston22}.  The age difference is caused by ongoing star formation in the disc after star formation has ended in the bulge.  Star formation is dependent on gas.  Gas could have been used up in the bulge before the disc due to such things like AGN activity (important in high mass galaxies \citep{croton06}).  Gas could also still be being accreted onto the disc after accretion has ended for the bulge.  Determining the exact scenario would require more and different data. The actual age difference in linear units is similar between light-weighted and mass-weighted measurements.

It is not clear why for light-weighted measurements galaxies with lower \Sersic\ index galaxies have greater difference between the age of their bulge and disc.  There is no obvious physical reason for this, and one might expect that galaxies with less defined bulges would have a closer age difference between bulge and disc.  Nonetheless, the result is quite significant, at $4.1 \sigma$ result.

It is possible that the age difference seen for specific star formation samples could be a reflection of the stellar mass difference seen for our sample.  However it could also be due to the fact that galaxies with large sSFR now are more likely to have the star formation in their discs.  This would decrease the stellar age of the disc relatative to the bulge giving rise to the age pattern seen here.  In particular this would explain the larger difference seen for the light-weighted measurement that traces the young, bright stars.  

Low-mass galaxies in our light-weighted sample have stellar age differences between bulge and disc close to zero, while higher mass galaxies have significant differences between bulge and disc ages.  This suggests that inside-out quenching for high mass galaxies is important, while smaller galaxies quench across the whole galaxy at the same time.  In the more massive galaxies AGN may be important in stopping the star formation in the bulge while leaving, for a while at least, star formation continuing in the disc.  In more massive systems, the light-weighted measurement shows a larger difference in bulge--disc age than the mass-weighted measurement, because the disc is able to continue forming stars long after the peak of star formation that the mass-weighted measurement tracks.   

\citet{frasermckelvie18} found that for low-mass lenticular galaxies the bulge is slightly younger than the disc, while for high-mass lenticular galaxies the bulge tends to be older than the disc.  \citet{pak21} and \cite{johnston22} also find similar trends with stellar mass for their lenticular samples.  \citet{dominguezsanchez20} measured stellar population gradients in MaNGA lenticular galaxies and found the more massive galaxies showed strong age gradients while the less massive ones showed relatively flat age gradients.  These results match our results for low- and high-stellar-mass samples, except the galaxies here are a mixture of late and early types.

Early-type galaxies have measured stellar ages differences between their bulge and disc close to zero. In early-type galaxies star formation has mostly stopped, and this result suggest that it stopped in the bulge and disc at the same time.  This result was also found by \citet{goddard17a,ferreras19,tabor19,barsanti20} for similar early-type samples. \citet{dominguezsanchez20} found for lenticular galaxies that the less massive ones showed relatively flat age gradients as did \citet{johnston22}. Our sample has been corrected for stellar mass completeness, so should be dominated by the low mass galaxies agreeing with these results. \citet{johnston12} and \citet{johnston14} actually find bulges younger than discs for lenticular galaxies going further than our result. This difference may be due to the fact their sample of galaxies are in clusters where ram pressure stripping is pushing the disk gas into the centre of a galaxy where it is used up in a final burst of star formation, making the bulge stellar population younger than the disc. Our sample does not include galaxies in such dense environments. A similar result is found by \citet{goddard17a} who find a positive age gradient in mass-weighted measurement for early-type galaxies not limited to clusters. \citet{santucci20} also find age gradients that are slightly positive in early type galaxies. However, \cite{zheng17} found in their mass-weighted measurements that early-type galaxies have negative age gradients. \citet{pak21}, in their lenticular sample, found that most bulges are older than discs. The differences here between papers are likely due to selection effects on the samples used, particularly on the definition of what constitutes an early-type galaxy with some gradient work including ellipticals and lenticular galaxies in the same sample.  In particular \cite{zheng17} defines their early type sample based on \Sersic\ index which significantly differs from our and other morphological selection.  

Late-type galaxies, in contrast to early-type galaxies, are found to have a much older bulge than disc in our results. Star formation has continued in the disc after the bulge has used up all its gas. \citet{goddard17a} found in their sample a negative age gradient in light-weighted measurements of late-type galaxies, supporting our result. However, in their mass-weighted measurement for late-type galaxies they find a near-zero age gradient. \cite{zheng17} did find a negative age gradients in their mass-weighted measurements of late-type galaxies in their sample. \citet{sanchezblazquez14} find shallow negative stellar age gradient in their late type sample for light weighted measurements but zero gradient for mass weighted measurements. \citet{delgado15} found in their sample, negative age gradients that steepen from E and S0 to Sbc and are shallower from Sbc to Sd.  There is some variation in the results presented here suggesting the differences in the method or selection of galaxies is playing a role.  The fact that we are comparing to gradients when we are not using them may be a significant factor.  Our mass-weighted measurement is a lot less than our light-weighed measurment for late type galaxies so could be be seen as near zero gradient when doing gradient measurements.  

There is no trend in bulge-disc age with environment as determined by the projected distance to the 5th nearest neighbour for the sample, as was found by \cite{johnston22}.  \citet{pak21} found that lenticular galaxies have larger bulge-disc age differences in higher-density environments.  The difference here can probably be explain with the different determination of environment with \citet{pak21} using projected stellar mass density while we use projected distance to the 5th nearest neighbour.

\citet{tacchella19} in the IllustrisTNG simulation find no significant metallicity difference between spheroid and disc at any stellar mass and argue that this must be due to efficient feedback that mixes the metals. This disagrees with what we find, that there are significant positive metallicity differences for all stellar masses.  Our result seems to imply that the feedback is not as strong as in the simulations. \citet{tacchella19} find, for mass-weighted measurements of simulated IllustrisTNG galaxies, that spheroids are older than discs, being log age \around 0.022 different at \around$\rm 11.5\log M_\odot$ and log age \around 0.15 different at \around $9\, \rm log M_\odot$.  For high stellar masses, our mass-weighted measurement agree with the IllustrisTNG result, but our low stellar masse measurement does not agree.  Here we find that the age difference is similiar in size to the high stellar mass measurement.  This likely indicates that disc formation occurred generally faster than the simulation predicts in smaller galaxies.  It should be noted that \citet{tacchella19} use the kinematics of a galaxy to select the bulge population which is significantly different from how we select the bulge population.  However they seem confident that photometric bulge-disc decompositions and kinematical estimates result in similar conclusions.

%------------------------------------------------------------------------------%

\section{Conclusion}
\label{conclusion}

The Mapping Nearby Galaxies at APO (MaNGA) survey is a large, optical, integral ﬁeld spectroscopic survey of low-redshift galaxies spanning a broad range in stellar mass, star formation rate, \Sersic\ index, and morphology. From this survey, we measure and compare the stellar populations of the bulge- and disc-dominated regions identified from their \Sersic\ profiles. Weighted medians of the metallicity and stellar age difference between bulge- and disc-dominated regions are measured, taking into account the completeness of the sample. 

For the entire sample, we find that generally bulge regions have higher metallicities than disc regions in both light-weighted and mass-weighted measurements. One likely cause of this difference is the greater gravitational potential within bulges compare to disc that enable them to retain more of their metals that are produced in star formation. Galaxies with higher stellar masses tend to have higher metallicity differences between bulge regions and disc regions than lower stellar mass galaxies, with light-weighted measurements showing the greatest difference. Again the great gravitational potential of bulges in galaxies with high stellar masses is the likely reason for this difference.  Early and late-type galaxies have similar metallicity differences between bulge regions and disc regions. There is no trend in metallicity difference between bulge regions and disc regions with environment, at least over the range that the MaNGA sample covers. Central galaxies tend to have greater metallicity difference between their bulge regions and disc regions than satellite galaxies at similar stellar mass. Influx of lower metallicity gas into the disc of centrals either by accretion or mergers could be the cause.  Quenching of satellite discs is the other possible explanation.

For the entire sample, we find that bulge regions generally have slightly higher stellar ages than disc regions in both light-weighted and mass-weighted measurements.  Ongoing star formation in the disc fuelled by more gas, possibly fed by accretion, is the likely cause.  AGN activity in the bulge, halting star formation there, may also be a factor.   Galaxies with higher specific star formation also have greater age differences between bulge and disc.  Our lower stellar mass sample shows little difference between bulge regions and disc regions stellar ages, while our higher stellar mass light-weighted sample has a median stellar age difference between bulge regions and disc regions that is positive. This suggests inside-out quenching for high-mass galaxies, while smaller galaxies quench across the whole galaxy simultaneously. Early-type galaxies tend to have bulge regions about the same stellar age as their disc regions, while late-type galaxies have a significantly positive difference between bulge regions and disc regions stellar age. We find no trend in stellar age difference between bulge regions and disc regions with environmental density for the range that MaNGA probes.  Central and satellite galaxies have similar stellar age differences between their bulge and disk regions.

The Milky Way has a metallicity difference between bulge and disc that lies close to the typical results for our sample, indicating that it is a normal galaxy in this respect. However, the Milky Way has a stellar age difference between bulge and disc that is significantly higher than the typical result we find here. This larger age difference may in part be due to the Milky Way measurement having better separation of the bulge and disc light than is possible in our measurements.

%------------------------------------------------------------------------------%

\section*{Acknowledgements}

%------------------------------------------------------------------------------%

Funding for the Sloan Digital Sky Survey IV has been provided by the Alfred P. Sloan Foundation, the U.S. Department of Energy Office of Science, and the Participating Institutions. SDSS acknowledges support and resources from the Center for High-Performance Computing at the University of Utah. The SDSS web site is www.sdss.org.

SDSS is managed by the Astrophysical Research Consortium for the Participating Institutions of the SDSS Collaboration including the Brazilian Participation Group, the Carnegie Institution for Science, Carnegie Mellon University, the Chilean Participation Group, the French Participation Group, Harvard-Smithsonian Center for Astrophysics, Instituto de Astrof\'isica de Canarias, The Johns Hopkins University, Kavli Institute for the Physics and Mathematics of the Universe (IPMU) / University of Tokyo, the Korean Participation Group, Lawrence Berkeley National Laboratory, Leibniz Institut f\"ur Astrophysik Potsdam (AIP), Max-Planck-Institut f\"ur Astronomie (MPIA Heidelberg), Max-Planck-Institut f\"ur Astrophysik (MPA Garching), Max-Planck-Institut f\"ur Extraterrestrische Physik (MPE), National Astronomical Observatories of China, New Mexico State University, New York University, University of Notre Dame, Observat\'orio Nacional / MCTI, The Ohio State University, Pennsylvania State University, Shanghai Astronomical Observatory, United Kingdom Participation Group, Universidad Nacional Aut\'onoma de M\'exico, University of Arizona, University of Colorado Boulder, University of Oxford, University of Portsmouth, University of Utah, University of Virginia, University of Washington, University of Wisconsin, Vanderbilt University, and Yale University.

Nicholas Scott acknowledges support of an Australian Research Council Discovery Early Career Research Award (project number DE190100375). Tania M.~Barone is supported by an Australian Government Research Training Program Scholarship. Francesco D'Eugenio acknowledges funding through the H2020 ERC Consolidator Grant 683184 and the ERC Advanced grant 695671 ``QUENCH'' and support by the Science and Technology Facilities Council (STFC).

We wish to thank Gary Da Costa, Ken Freeman and Mike Bessell for their valuable insight. We wish to thank Christopher Delfs for his valuable assistance. We wish to thank Anne-Marie Weijmans, Mariangela Bernardi, Benjamin Alan Weaver Rodrigo Tobar Carrizo and Mina Pak for their valuable help.

%------------------------------------------------------------------------------%

\bibliography{MaNGA_plah}

%------------------------------------------------------------------------------%

\section*{Appendix}

\label{Appendix}

This Appendix gives a detailed summary of the conclusions reached by the papers listed in the Introduction.

There is extensive literature showing that the inner regions of galaxies have more metals than outer regions.
\begin{itemize}

\item \citet{moorthy06} observed 38 galaxies with well-defined bulges and discs using a long-slit spectrograph. They found that most bulges have decreasing [Z/H] with increasing radius, and that galaxies with larger central metallicities have steeper gradients.

\item \citet{macarthur09} examined eight star-forming, spiral galaxies using long-slit observations and found a small decrease with radius in [Z/H] for at least four of their sample.

\item \citet{sanchezblazquez11} found, in long-slit observations of 4 galaxies ranging from lenticulars to late-type spirals, a decrease in [Z/H] in three of the four galaxies.

\item \citet{johnston12} used long-slit observations of nine lenticular galaxies (S0s) in the Fornax cluster to reveal that, where the bulge and disc could be distinguished, the bulges have systematically higher metallicities.

\item In \citet{johnston14}, a further 21 lenticular galaxies in the Virgo cluster were examined using a long-slit spectrograph, and the bulges were found to be more metal-rich than their surrounding discs.
  
\item \citet{ganda07} observed 18 late-type spiral galaxies with an integral-field spectrograph and showed that [Z/H] decreases from the bulge to the disc.

\item IFU observations by \citet{sanchezblazquez14} of 62 spiral galaxies found mean metallicity gradients are shallow and negative.
  
\item \citet{delgado15} observed 300 galaxies with integral field spectroscopy for a wide range of Hubble types as part of the CALIFA survey. In their sample [Z/H] gradients are mildly decreasing profiles for most Hubble types except Sd galaxies, which show little decrease; most of their galaxies have stellar masses larger than 10$^{10}$M$_{\odot}$.

\item \citet{goddard17a} reported on observations of 721 MaNGA galaxies showing that, in light-weighted [Z/H] measurements, early-type galaxies have a negative gradient.

\item \citet{zheng17} used MaNGA observations of 1005 galaxies to examine galaxies of all types from the  survey. They find, in mass-weighted measurements, that the mean [Z/H] gradients are close to zero but slightly negative for both early and late-type galaxies.

\item \citet{frasermckelvie18} studied 279 lenticular galaxies from the MaNGA survey and found that the higher mass lenticulars have bulges more metal-rich than their discs, though in general bulges and discs had similar metallicities.

\item \citet{tabor19} studied 273 early-type galaxies from the MaNGA survey and found that bulges have higher metallicites than discs, which tend to span a wider [Z/H] range indicating a more varied star formation history compared to bulges.

\item In a sample of 96 passive central galaxies with integral-field spectroscopy from the SAMI Galaxy Survey measured out to \around 2R$\rm _e$, \citet{santucci20} found negative [Z/H] radial gradients and evidence that gradients become shallower with increasing stellar mass. They also found that metallicity gradients for fixed stellar mass were similar for satellites and central galaxies.

\item \citet{barsanti20} studied 192 lenticular galaxies in the cluster sample of the SAMI Galaxy Survey using 2D photometric bulge-disk decomposition and found that most galaxies have more metals in the bulge than disc.

\item \citet{zibetti20} looked at 69 early-type galaxies from the CALIFA integral field spectroscopic survey. They found that  early-type galaxies universally had strong, negative metallicity gradients within 1~R$\rm _e$ that flatten out at larger radii.
  
\item \citet{johnston22} found in 78 lenticular galaxies in the MaNGA sample that bulges are generally more metal rich than discs.

\end{itemize}

However, the literature is not unanimous in finding more metals in the central regions of galaxies.
\begin{itemize}

\item \citet{goddard17a} found that both light-weighted and mass-weighted measurements of late-type galaxies showed near zero [Z/H] gradient, and mass-weighted measurements of [Z/H] gradients in early-type galaxies were also near zero.

\item \citet{ferreras19} examined 522 early-type galaxies from the SAMI Galaxy Survey, looking at the relationships between [Z/H], [Mg/Fe], [C/Fe] and age gradients compared with velocity dispersion, stellar mass, dynamical mass, surface stellar mass density, stellar potential and virial test ($\sigma^2$/R). They found a near zero [Z/H] gradient in their sample of galaxies.

\item \citet{dominguezsanchez20} measured stellar population gradients in MaNGA lenticular galaxies by stacking in bins of luminosity and central velocity dispersion. Galaxies with masses above $\rm 3 \times 10^{10} M_\odot$ show little or no metallicity gradients, while less massive ones show a significant metallicity gradient. 
  
\item \citet{pak21} used photometric bulge-disk decomposition to separate bulges and discs for 34 lenticular galaxies from the CALIFA survey. In their sample they find no no systematic difference in metalllicity between bulges and their associated discs. They also found that the bulge-disc metallicity difference does not vary as a function of local density.
  
\end{itemize}

The literature includes many studies showing that the stars in the inner regions of galaxies tend to be older.

\begin{itemize}

\item \citet{sanchezblazquez14} showed that the mean stellar age gradient has a shallow but negative slope for their sample of spiral galaxies.

\item \citet{delgado15} found in their sample, negative age gradients steepen from E and S0 to Sbc and are shallower from Sbc to Sd.  Milky Way-like (Sbc) galaxies thus stand out as those with the steepest radial age profiles.

\item \citet{goddard17a} found a negative age gradient in lighted-weighted measurements of late-type galaxies.

\item \cite{zheng17} found that mass-weighted measurements yield negative age gradients for both early-type and late-type galaxies.

\item \citet{frasermckelvie18} found that, for massive lenticular galaxies, the bulge was older than the disc, although for most of their sample they found that bulge and disc ages were similar.

\item In the lenticular sample of \citet{dominguezsanchez20}, the more massive galaxies showed strong age gradients, while the less massive ones showed relatively flat age gradients.

\item In their CALIFA galaxy sample of lenticular galaxies, \citet{pak21} found that most bulges are older than their associated discs.  They also find that for lenticular galaxies the difference between their bulge and disc age increase with increasing stellar mass.  Additionally they find that lenticular galaxies have larger age differences in higher-density environments.  

\item \citet{johnston22} found in lenticular galaxies that bulges are generally older than discs.  They also found no clear difference in the formation or quenching processes comparing bulges and discs as a function of galaxy environment.  

\end{itemize}

Some sources in the literature have found that the bulge and disc have similar stellar ages.
\begin{itemize} 

\item \citet{sanchezblazquez11} found a fairly flat age gradient in their four galaxy sample.

\item The mass-weighted measurements of \citet{sanchezblazquez14} for late-type galaxies show zero stellar age gradient.

\item \citet{goddard17a} found a near zero age gradient in the light-weighted measurements of early-type galaxies and mass-weighted late-type galaxies in their sample.

\item \citet{ferreras19} also found near zero age gradient in early-type galaxies from the SAMI Galaxy Survey.

\item One example is the measurements made by \citet{tabor19} of early-type galaxies from the MaNGA survey.

\item \citet{zibetti20} in the CALIFA survey found age profiles that have similar ages at the centre and the outskirts of the galaxies, but are U-shaped in between, with a minimum around 0.4~R$\rm _e$ and asymptoting to a maximum age beyond 1.5~R$\rm _e$. 

\item \citet{barsanti20} found in their cluster, lenticular sample that most galaxies have bulges and disks with statistically indistinguishable ages, though there are fair numbers of galaxies with bulge ages both older and younger than their associated discs.

\item \citet{johnston22} found in low mass lenticular galaxies ($\rm \le 10^{10} M_{\odot}$) that bulges and discs tend to have similar ages.

\end{itemize}

A few sources in the literature show inner region younger than outer regions.
\begin{itemize} 

\item The \citet{johnston12} lenticular galaxy sample showed younger stellar populations in their bulges than the discs, as did the results of \citet{johnston14}. Both of these samples were in clusters, and they suggested that ram pressure stripping is pushing the disc gas into the centre of the galaxies where it is used up in a final burst of star formation, making the bulge stellar population on average younger than the disc.

\item \citet{goddard17a} found in their sample of early-type galaxies that the mass-weighted stellar age showed an increasing gradient.

\item \citet{frasermckelvie18} found that lower-mass lenticulars tend to have bulges marginally younger than their discs.

\item In a passive central galaxy sample from the SAMI Galaxy Survey, \citet{santucci20} find age gradients that are slightly positive and no difference in the trend between central and satellite galaxies.

\end{itemize}

%------------------------------------------------------------------------------%

\end{document}